\documentclass[runningheads]{llncs}
\usepackage[T1]{fontenc}
\usepackage{tabularx}
\usepackage{graphicx}
\usepackage{amsmath}
\usepackage{hyperref}
\usepackage{amsmath}
\usepackage{algorithm}
\usepackage{algorithmicx}
\usepackage{algpseudocode}
\usepackage{subcaption} 
\usepackage{orcidlink}

\usepackage{color}

\begin{document}
\title{A BERT Based Hybrid Recommendation System For Academic Collaboration\thanks{Vellore Institute of Technology, Chennai, India}}
\titlerunning{Hybrid Recommendation System for Academia}
%
\author{Sangeetha N\inst{1}\orcidlink{0000-0001-5320-817X} \and
Harish Thangaraj\inst{2}\orcidlink{0009-0007-7364-4797} \and
Varun Vashisht\inst{3}\orcidlink{0009-0002-3466-4676} \and
Eshaan Joshi\inst{4}\orcidlink{0009-0005-7527-2283}\and
Kanishka Verma\inst{5}\orcidlink{0009-0000-6862-5670}\and
Diya Katariya\inst{6}\orcidlink{0009-0006-2640-7732}}

\authorrunning{Sangeetha N et al.}
%
\institute{Vellore Institute of Technology, Chennai, India \\
\email{n.sangeetha@vit.ac.in} \inst{1} \\
\email{harish.thangaraj03@outlook.com} \inst{2}\\
\email{varunvashisht.work@gmail.com} \inst{3}\\
\email{eshaanjoshi.work@gmail.com} \inst{4}\\
\email{kanishkaverma2003@gmail.com} \inst{5}\\
\email{diyakatariya0907@gmail.com} \inst{6}
\\
}

\maketitle              
\begin{abstract}
Universities serve as a hub for academic collaboration, promoting the exchange of diverse ideas and perspectives among students and faculty through interdisciplinary dialogue. However, as universities expand in size, conventional networking approaches via student chapters, class groups, and faculty committees become cumbersome. To address this challenge, an academia-specific profile recommendation system is proposed to connect like-minded stakeholders within any university community. This study evaluates three techniques: Term Frequency-Inverse Document Frequency (TF-IDF), Bidirectional Encoder Representations from Transformers (BERT), and a hybrid approach to generate effective recommendations. Due to the unlabelled nature of the dataset, Affinity Propagation cluster-based relabelling is performed to understand the grouping of similar profiles. The hybrid model demonstrated superior performance, evidenced by its similarity score, Silhouette score, Davies-Bouldin index, and Normalized Discounted Cumulative Gain (NDCG), achieving an optimal balance between diversity and relevance in recommendations. Furthermore, the optimal model has been implemented as a mobile application, which dynamically suggests relevant profiles based on users' skills and collaboration interests, incorporating contextual understanding. The potential impact of this application is significant, as it promises to enhance networking opportunities within large academic institutions through the deployment of intelligent recommendation systems.

\keywords{academic networking \and university collaboration \and hybrid recommendation system \and TF-IDF \and cosine similarity \and BERT \and affinity propagation \and NDCG}
\end{abstract}
\section{Introduction}
Networking is a fundamental skill crucial for both professional advancement and personal development. In the dynamic landscape of modern university environments, fostering intellectual exploration and academic collaboration is paramount. However, as institutions grow in size and complexity, the task of meaningful interaction becomes increasingly challenging. Consider a newly enrolled student navigating a large university campus, eagerly seeking opportunities to collaborate on projects and connect with like-minded individuals. The traditional methods of networking such as approaching student groups and seeking guidance from professors though time-honored, often prove to be cumbersome and inefficient. These methods demand significant time and energy, leading to delays in the execution of several ideas.

This article aims to address the inherent limitations and inefficiencies of traditional networking methods within a university setting by proposing a digital networking solution operating in academic circles. Enter ‘FindMate’ – a swipe-based recommendation system presented in the form of a mobile application. While TF-IDF has been a popular approach to build such systems due to its simplicity and effectiveness in text representation, BERT \cite{BERT} with its deep contextual embeddings and superior understanding of language nuances offers better capturing of semantic meaning and thereby better recommendations. This paper proposes a system leveraging the best of both techniques for our use case, a hybrid model that uses both TF-IDF and BERT embeddings. The study assesses all three techniques on custom generated profile data. The results provide a clear evaluation on each of the model's capability to group variety of cross-domain data. The recommendation approach with the best metrics is integrated into the mobile app revolutionizing the concept of campus networking with contextual awareness and language understanding. 

Despite the widespread adoption of recommendation systems in academic circles, the existing literature predominantly focuses on facilitating access to research papers, certain academic resources and there is limited work on integrating language models like BERT for the same. However, this study endeavours to extend the scope of academic recommendation systems to include the realm of profile connections within university environments using the mentioned hybrid approach. ‘FindMate’ aims to bridge the gap between individuals with similar interests and aspirations, thereby fostering collaborative partnerships and catalyzing innovative projects.

\section{Related Works}

\subsection{Social Media for Networking} The web is full of networking solutions to facilitate connections for academia. Several existing platforms and research studies have explored similar objectives, aiming to address the challenges associated with traditional networking. From approaching peers in person to connecting on digital platforms, there have been significant advancements in networking methodologies. Udenze and Silas \cite{ref1} in their paper study academic awareness and the usage of LinkedIn for academic networking using a non-probability-based sampling method and conclude that scholars have not fully embraced the power of LinkedIn to build academic collaboration due to the low level of awareness. Similarly, Ritesh Chugh et al \cite{ref11} conducted a scoping survey review focusing on the use of social media by academics highlighting the need for higher education institutions to provide awareness to its stakeholders of the applicability of social media for academia. A study on the benefits and problems of using general social networking sites in the scholastic domain has concluded to be a lot more negative than positive as observed by Katy Jordan and Martin Weller \cite{ref2} in their research. Generic social networking platforms prioritize commercial usage and marketing rather than upholding academic interests. On a more fundamental side, Heffernan \cite{ref12} studies consequences of academic networking and surveys the effect on overall career opportunities and faculty relationships. His findings suggest that merit-based achievements can be overshadowed by network opportunities and good connections impact individual's aspirations proving to be a key motivation to our work.

\subsection{Academic Specific Networking and Recommendation Systems} Conole et al \cite{ref13} present 'Cloudworks' a web application developed to share learning and teaching ideas across users. The work also explores various theoretical frameworks towards connecting people and the downside of maintaining a self sustaining user base for such platforms. 
Adeniyi et al \cite{ref3} in their paper, investigate the significance of academic-specific social networking tools in contrast to the generic platforms discussed previously. The study highlights the poor usability of such niche-specific platforms and proves to be clumsy for a new user, spending a significant amount of time getting acquainted with the interface. Further, academic sites lack a personalized touch of giving you relevant collaboration information due to the sheer volume of data as concluded by Kong et al \cite{ref4} in their publication. Ko et al \cite{ref16} review recommendation models and techniques giving comprehensive overview of existing systems in application service fields further enhancing the basic understanding of such systems for this study . Zhang et al \cite{ref6} investigate various scholarly recommendation systems including literature and author collaborator systems emphasising the importance for better techniques than content based and collaborative filtering. Nikhat Akhtar and Devendra Agarwal \cite{ref7} explore a research paper recommendation system employing several machine learning paradigms . Similarly, a program and course recommendation algorithm was discussed by Mohammed Ezz and Ayman Elshenawy \cite{ref8} in their research . StudieMe is a college suggestion system published by Vidish Sharma et al \cite{ref9} that uses text-based Cosine Similarity to match relevant data. Presenting a collaborative filtering recommendation system, Jianjun Ni et al \cite{ref15} use a two step approach applying TF-IDF and fuzzy logic to recommend data. 

\subsection{Deep Learning Based Recommendation Systems} The previous references primarily focused on conventional methods for building recommendation systems. Shifting our literature survery towards deep learning architectures, Rodríguez-Hernández et al \cite{ref17} conducted an experimental study where they compare traditional recommendation algorithms with BERT based ones and conclude that while BERT techniques have been used well in Natural Language Processing tasks the exploitation of BERT in recommendation systems is a sparsely unexplored topic. Juarto and Girsang \cite{ref19} in their work present a neural system integrating Sentence BERT with a collaborative news recommender. This integration results in a significant increase in hit ratio indicating the positive impact using deep architectures. A text recommendation system fused with Convoluted neural networks (CNN) and BERT semantic information was offered by Xingyun Xie et al \cite{ref20} demonstrating an improved feature extraction mechanism when performing small scale Top-N recommendations.  

\section{The Proposed Solution and Implementation}
The development of the solution consists of 5 broad segments.

\subsection{Initial Data Extraction and Pre-Processing}
\begin{figure}
\centering
\includegraphics[width=0.6\linewidth]{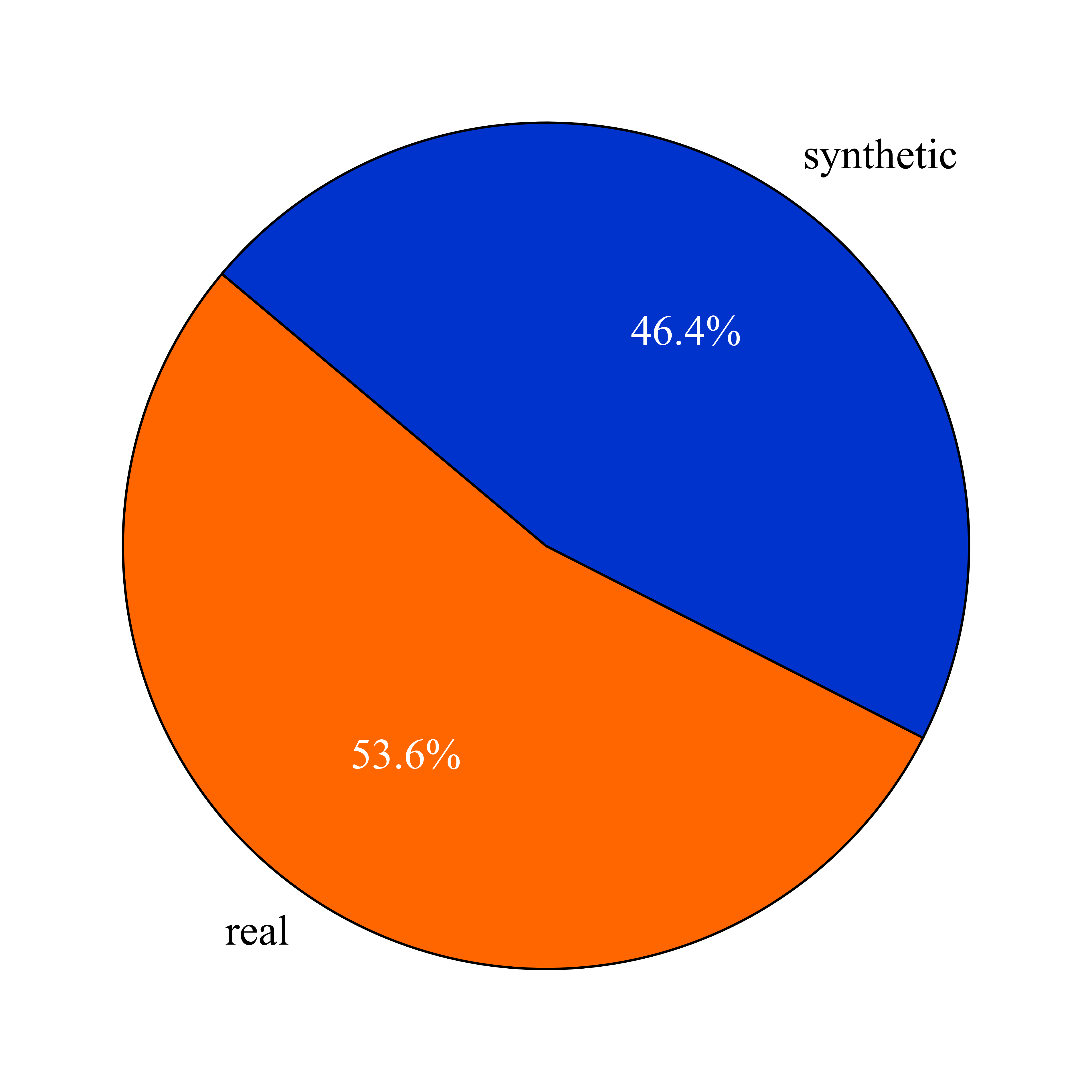} 
\caption{Distribution of synthetic vs real data}
\label{fig:1}
\end{figure}

The foundational text data was collected through a survey designed to gather user information, including experience, collaboration interests, and skills, as well as credentials such as name and email. To augment the dataset, we generated additional synthetic data for the same fields by defining a manual pool of skill set and domain interest, thereby creating a large and diverse corpus for comprehensive model training and testing (refer Fig.\ref{fig:1} for proportion). Following data collection, the dataset was thoroughly pre-processed using the Natural Language Toolkit (NLTK) and loaded onto MongoDB for real time storage and retrieval.

\begin{table}[h]
\centering 
\caption{Structural view of the survey dataset}
\label{tab:1}       
\begin{tabular}{lll}
\hline\noalign{\smallskip}
Column & Data type & Description  \\
\noalign{\smallskip}\hline\noalign{\smallskip}
Name & String & Full name of the user \\ 
Email & String & Email of the user \\
Profession & String & Profession of the user \\
Experience & number & Experience in years \\
Interest & String & Type of academic activity \\
Collaboration with & String & Preferred collaborators \\
Domain & String & Domain of collaboration \\
Skillset & String & Skillset of the user \\
\noalign{\smallskip}\hline
\end{tabular}
\end{table}

\begin{algorithm}
\caption{Text Vectorization and Similarity Evaluation}
\label{alg:1}
\begin{algorithmic}[1]
    \Require Corpus\_Document, Profile\_Data, Stopword\_Corpus, Technique
    \Ensure Similarity\_Score

    \State Load Corpus\_Document, remove stop words using Stopword\_Corpus
    
    \If{Technique = "TF-IDF"} 
        \State Apply stemming 
        \State Compute TF and IDF, vectorize with TF-IDF 
    \ElsIf{Technique = "BERT"} 
        \State Tokenize and generate BERT embeddings 
    \ElsIf{Technique = "Hybrid"} 
        \State Generate BERT embeddings 
        \State Compute TF-IDF, combine vectors 
    \EndIf

    \State Calculate Cosine Similarity between vectors
    \State Affinity Propagation clustering
\end{algorithmic}
\end{algorithm}

\subsection{TF-IDF}
TF-IDF and Cosine Similarity are established methodologies to build recommendation algorithms. TF represents the frequency of a term relative to the total word count. IDF reflects the proportion of documents in the
corpus that contain the term. TF-IDF together specify
the relevancy of a specific term in the given corpus in
the form of a numerical vector. Columns ‘domain’ and
‘skillset’ are combined and vectorized using the mentioned TF-IDF technique to grab relevant information
on the user’s skills and interests.

\begin{equation}
TF = \frac{\text{number \ of \ times \ the \ term \ appears}}{\text{total \ number \ of \ terms \ in \ the \ document}}
\end{equation}

\begin{equation}
IDF = \log \left( \frac{\text{number \ of \ documents}}{\text{document \ frequency \ of \ the \ term}} \right)
\end{equation}

\begin{equation}
TF \ IDF = \text{TF} \times \text{IDF}
\end{equation}

Subsequently, Cosine Similarity is applied to these
vectors to quantify the similarity between them. It calculates the cosine of the angle between the two vectors in a multidimensional space, signifying their orientations. Documents pointing in similar directions will have similar content, yielding a higher cosine similarity score which acts as the fundamental criteria to filter out relevant user profiles.

\begin{equation}
\text{similarity}(A, B) = \cos(\theta) = \frac{A \cdot B}{\|A\| \|B\|}
\end{equation}

The pre-processed data points were vectorized to compute the cosine similarity matrix, (workflow given by Algorithm.\ref{alg:1}). Given the unsupervised variable nature of the text, similar profiles
were grouped together using Affinity Propagation, a clustering method that automatically determines the optimal number of clusters operating on the calculated cosine similarity matrix. The resulting high dimension groups were reduced using t-distributed Stochastic Neighbor Embeddings (t-SNE) and plotted for visualization.

All profiles in the dataset were labeled (cluster based
relabeling) with their respective cluster numbers for
further evaluation and bench-marking of cohesion which
is discussed under results.

\subsection{BERT}
BERT’s architecture comprises transformer encoders that process input text bidirectionally, capturing contextual relationships between words. As in the case of profile matching, nuanced meanings and relationships between the sequences are grabbed effectively.
The input data is tokenized using a Masked and Permuted Pre-training Network (MPNet) Tokenizer \cite{Mpnet}. The pre-trained ’bert-large-uncased’ architecture \cite{BERT} is fine tuned on the input sequence to facilitate the downstream recommendation task. The model tokenizer generates embeddings for each profile, which are then used to compute cosine similarity scores. These scores measure the similarity between different profiles, analogous to the TF-IDF approach but with a deeper contextual understanding.
Affinity Propagation once again is used to cluster the alike profile embeddings which is then visualized, analyzed and benchmarked for comparison.

\subsection{Hybrid Approach}
The hybrid approach integrates both TF-IDF vectors and BERT embeddings to capture both term significance and the contextual nuances of the data points. To achieve this, the matrix representations generated in the previous experiments were combined with simple averaging. The similarity matrix was calculated on the combined matrix and was subjected to clustering.

\begin{figure}[h]
\centering
\begin{subfigure}{0.5\textwidth}
    \centering
    \includegraphics[width=\linewidth]{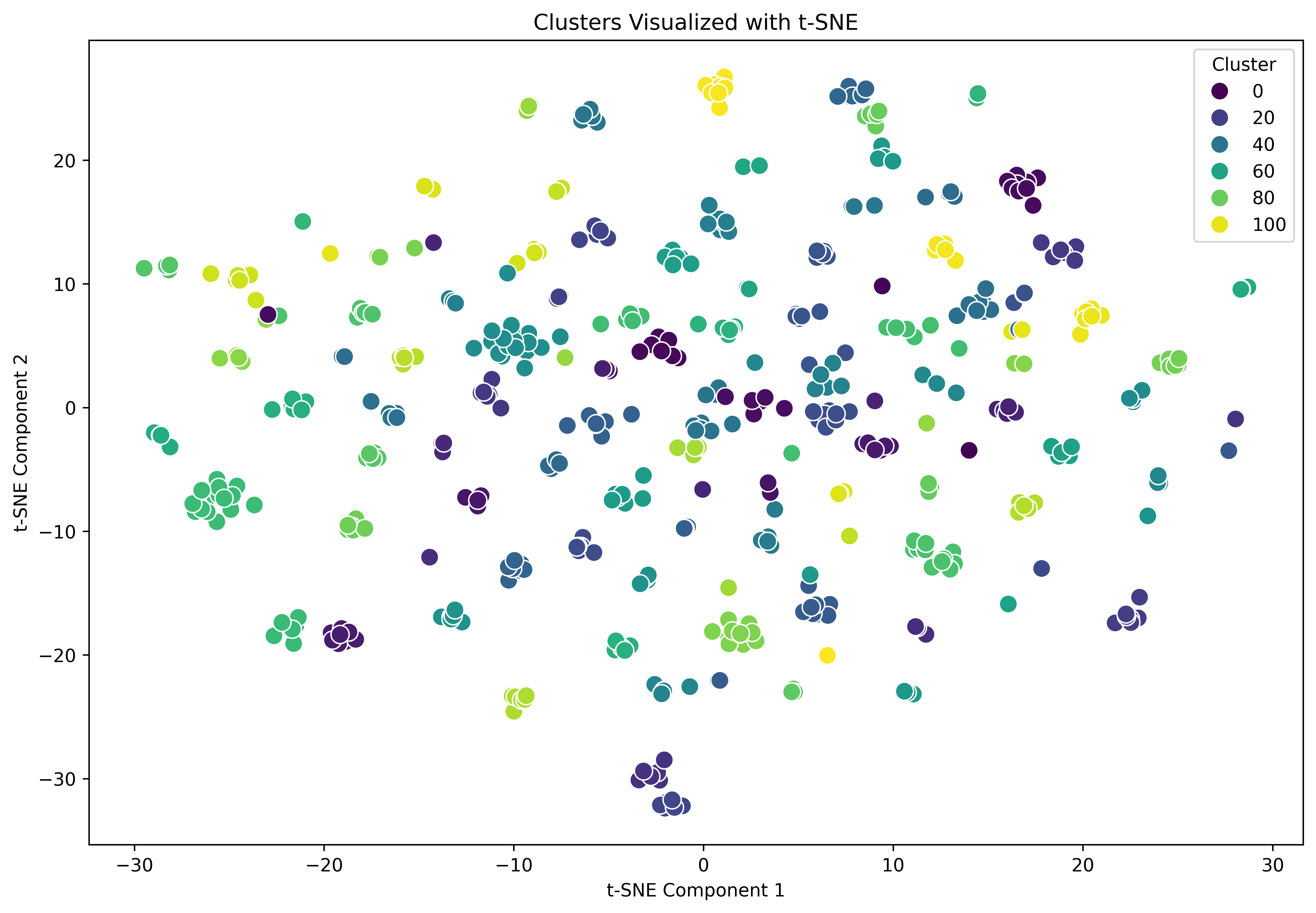}
    \caption{TF-IDF}
    \label{fig:3}
\end{subfigure}%
\hfill
\begin{subfigure}{0.5\textwidth}
    \centering
    \includegraphics[width=\linewidth]{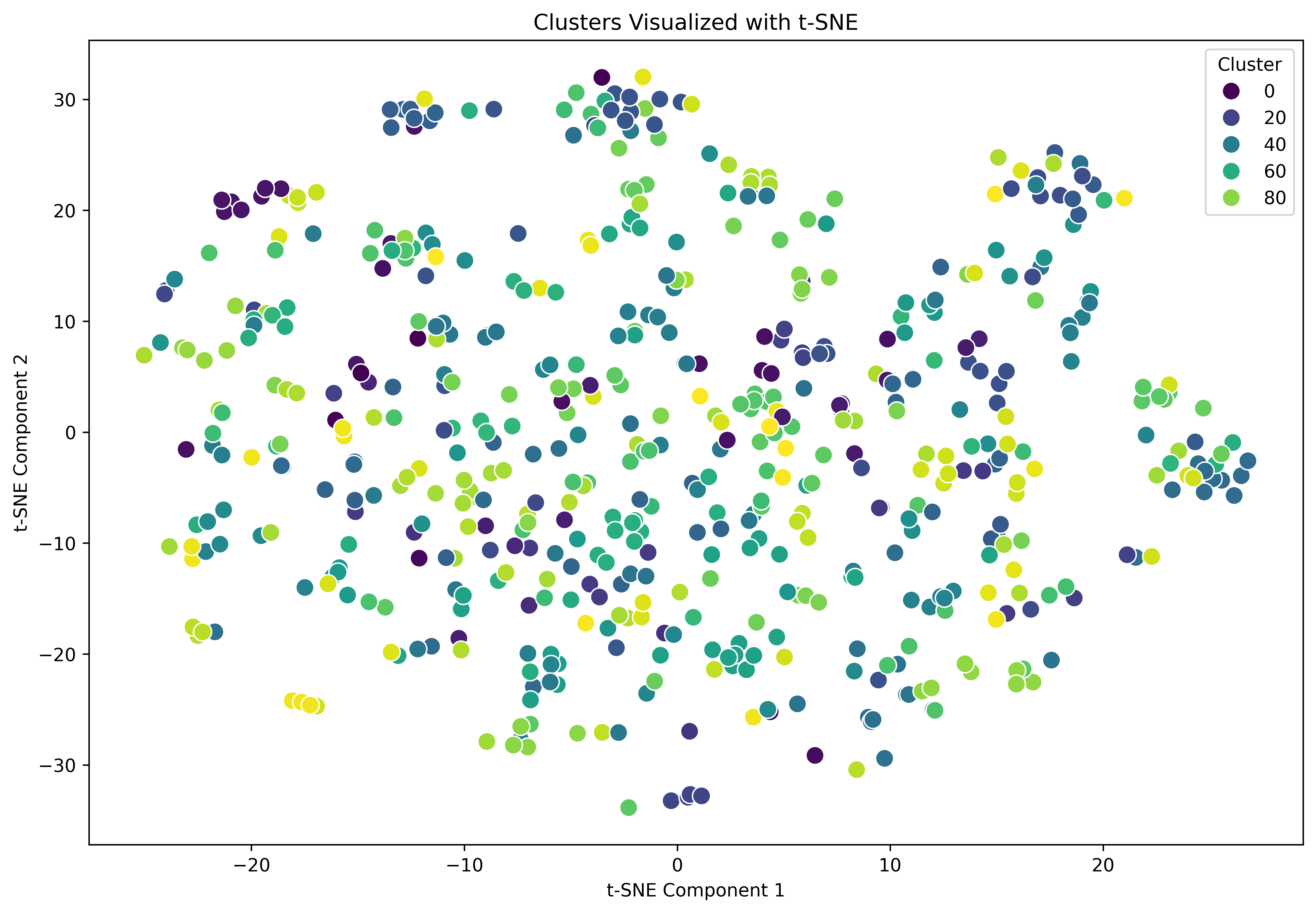}
    \caption{BERT}
    \label{fig:4}
\end{subfigure}

\vspace{1em} 

\begin{subfigure}{0.5\textwidth}
    \centering
    \includegraphics[width=\linewidth]{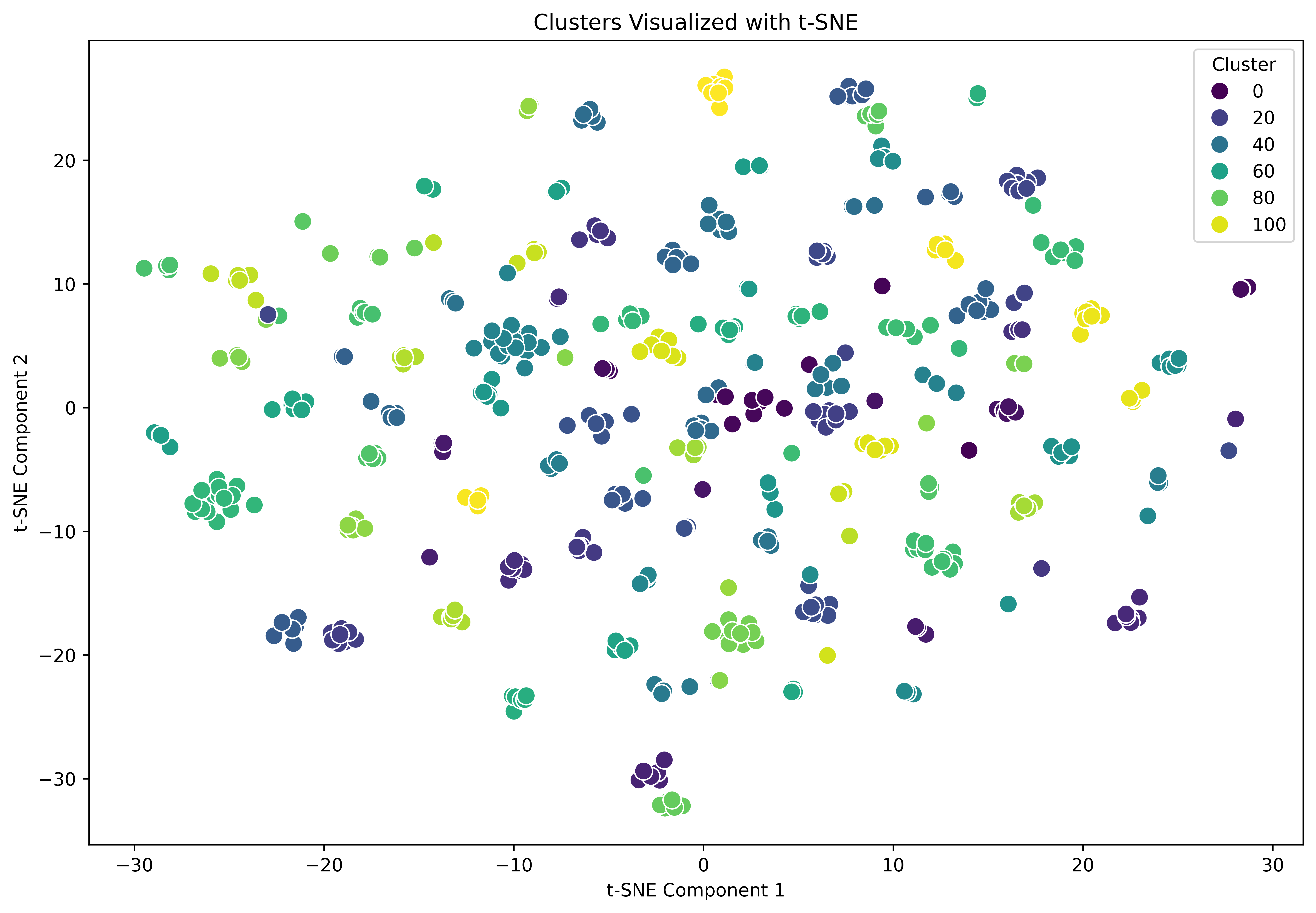}
    \caption{Hybrid}
    \label{fig:5}
\end{subfigure}
\caption{Affinity Propagation clustering}
\label{fig:clustering_comparison}
\end{figure}

Figures \ref{fig:3}, \ref{fig:4}, and \ref{fig:5} illustrate the clustering behaviour of the three methodologies examined. Based on these visualizations, the corresponding metrics and analysis are discussed in detail under the results section.

\subsection{Recommendation Engine and Mobile Application}
The algorithm has been integrated onto a mobile application for on the go usage. Upon user login, the system designates the logged-in user as the target user and generates recommendations in real-time. The columns ‘Interest’, ‘Profession’ and ‘Collaboration’ are used as filtering criteria for profile matches. For instance, if a user expresses a desire to collaborate with a faculty member, this preference is taken into consideration, faculties with high similarity scores are recommended based on their alignment with the user’s preferences.

\begin{figure*}[h]
  \centering
 \includegraphics[width=1.0\textwidth]{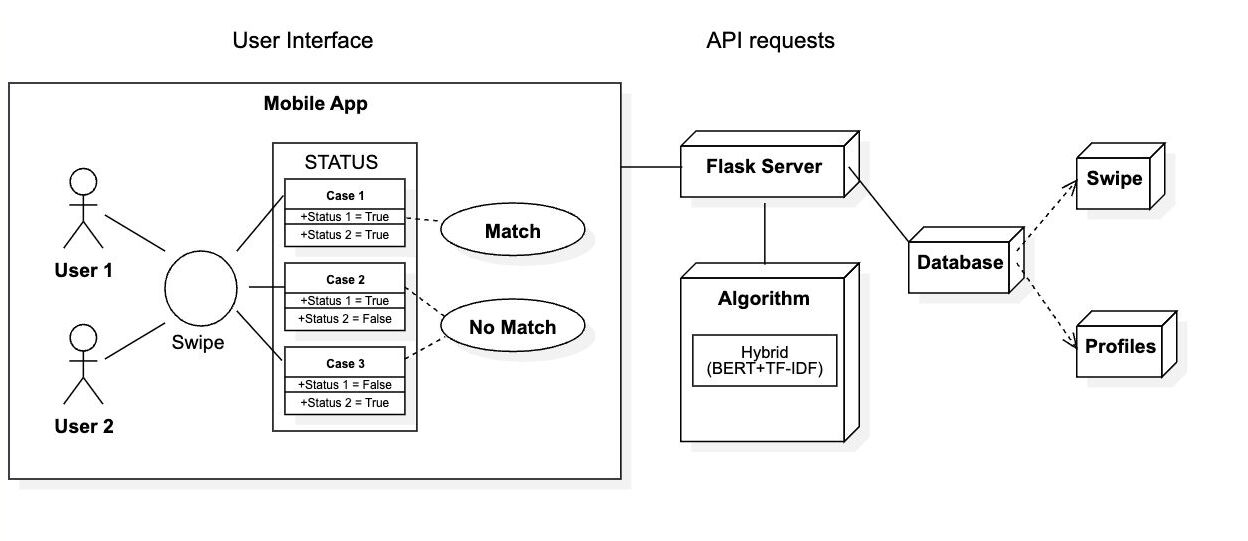}
  \caption{Solution architecture of the integrated system}
  \label{fig:6}
\end{figure*}

A user-friendly swipe-based matching feature is also implemented on the user interface. Two Boolean status variables keep track of the swipe actions. “Status 1” denotes the swipe action of “User 1” on “User 2”, while “Status 2” represents the swipe action of “User 2” on “User 1”. A match is established for further communication only when both users swipe right and a P2P chat window is enabled to further communication, indicating mutual interest to collaborate as depicted in Fig.\ref{fig:6}.

\begin{figure}[h!]
    \centering
    \begin{minipage}[b]{0.3\textwidth}
        \centering
        \includegraphics[width=\textwidth]{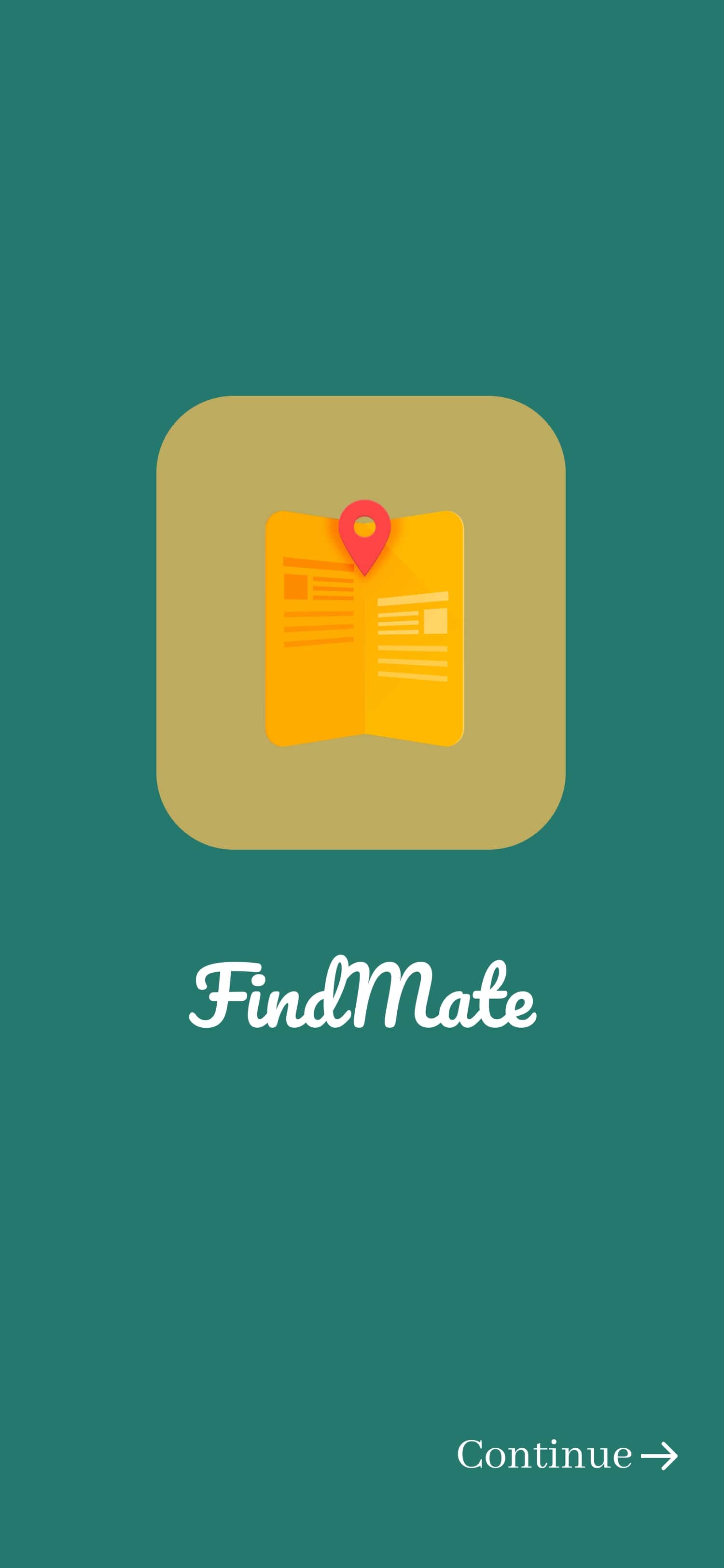}
        \caption{Home page}
        \label{fig:7}
    \end{minipage}
    \begin{minipage}[b]{0.3\textwidth}
        \centering
        \includegraphics[width=\textwidth]{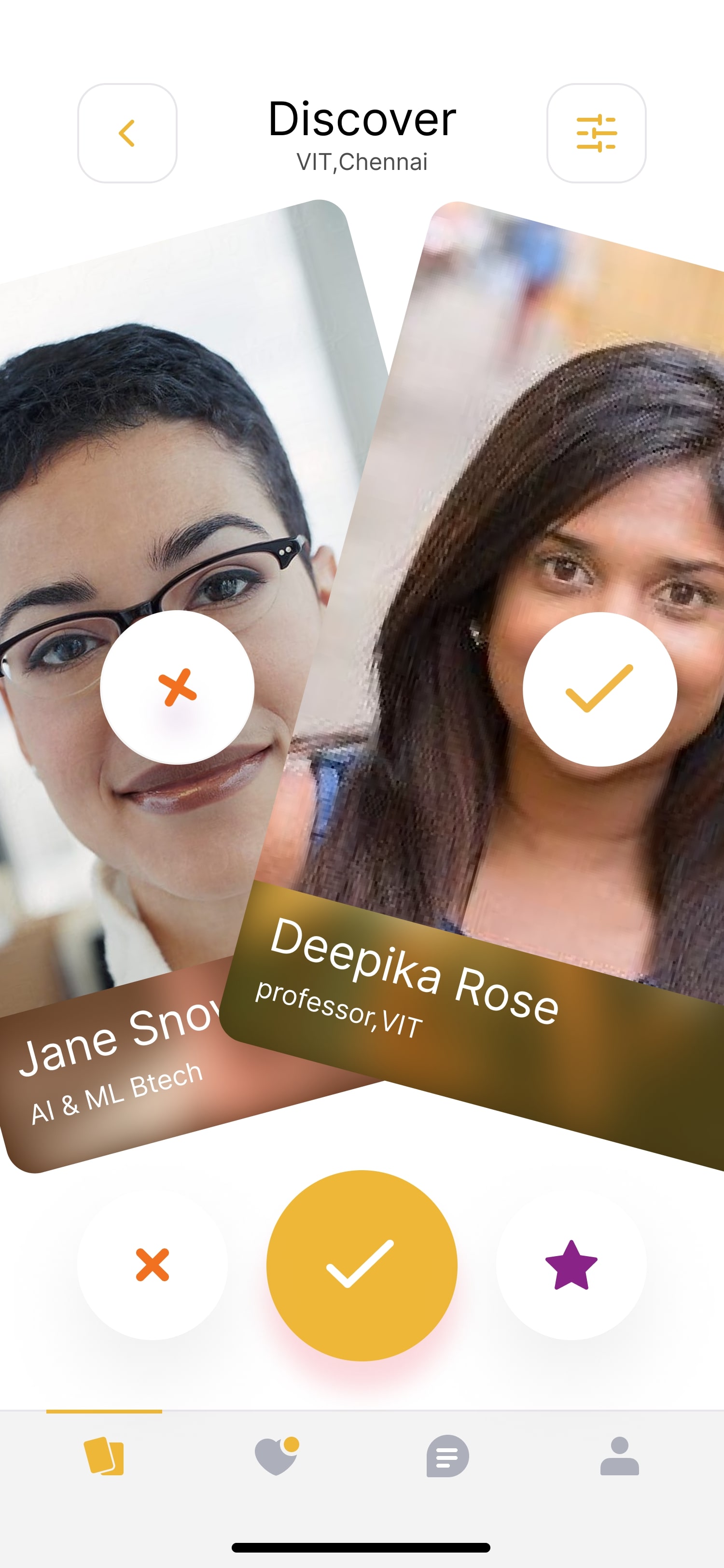}
        \caption{Swipe UI}
        \label{fig:8}
    \end{minipage}
    \begin{minipage}[b]{0.3\textwidth}
        \centering
        \includegraphics[width=\textwidth]{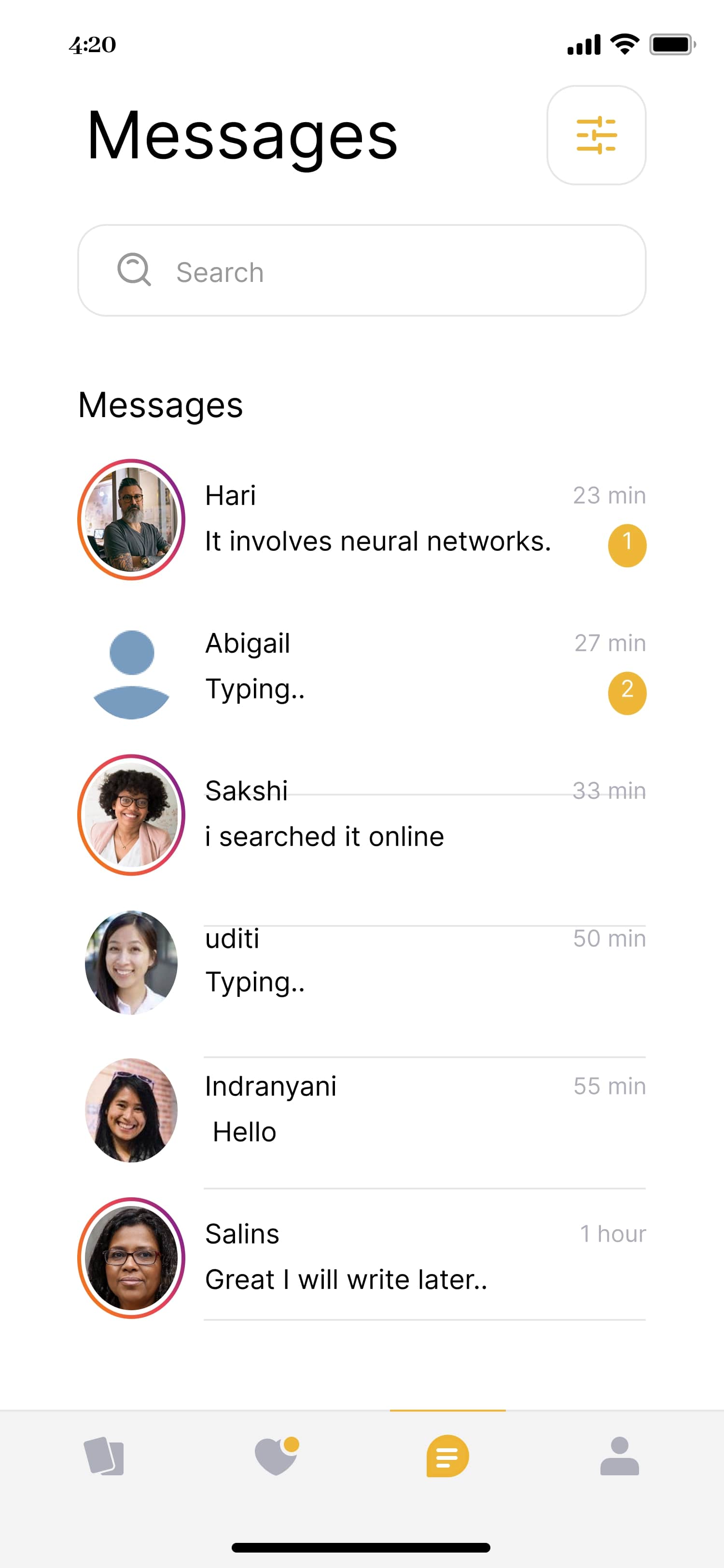}
        \caption{P2P chat feature}
        \label{fig:9}
    \end{minipage}
\end{figure}

To enhance security, sensitive information such as passwords are hashed before storage. To maintain a credible user base, a rating system is implemented where each profile is tagged with an average rating provided by other collaborators, allowing users to make informed decisions when selecting potential matches based on their rating. The application UI is presented in Fig.\ref{fig:7}, \ref{fig:8}, \ref{fig:9}.

\section{Results and Inference}
The evaluation for the three experiments presents computing intra-cluster similarity, Silhouette and Davies-Bouldin Index scores. The target fields are randomly distributed, and certain profiles may belong to multiple clusters. Therefore, these scores should be evaluated relatively rather than ideally, focusing on how model behavior changes across different implementations. Further, the models were also evaluated on two absolute metrics, namely NDCG and mAP (refer Table \ref{tab:2}). Let's first clarify the significance of these metrics for the data source. 

Intra-Cluster Similarity:
This metric quantifies the degree of similarity between data points within the same cluster. High intra-cluster similarity means that the points within a cluster are very similar to each other.

Silhouette Score:
The Silhouette Score evaluates the quality of clustering by measuring how well each data point is matched to its own cluster compared to other clusters. It ranges from -1 to 1, where a score close to 1 indicates that the data points are well-clustered, meaning they are more similar to their own cluster than to neighboring clusters.

Davies-Bouldin Index:
The Davies-Bouldin Index assesses the average similarity ratio of each cluster with its most similar cluster. Lower values of this index indicate better clustering results, as they reflect clusters that are both compact (internally cohesive) and well-separated from one another. 

NDCG: 
Normalized Discounted Cumulative Gain (NDCG) accounts for the position of relevant profiles (higher similarity scores) in the ranked list, assigning higher weights to relevant profiles that appear earlier. This provides a nuanced understanding of both relevance and ranking quality.

mAP:
The average precision for a single target profile is derived from the precision-recall curve, where precision is computed at each relevant profile retrieval based on similarity scores. By averaging these precision scores for all target profiles, mAP provides a comprehensive measure of the system's ability to recommend relevant profiles, ensuring that the most similar and relevant profiles are prioritized. 

\noindent

\begin{equation}
    \text{mAP} = \frac{1}{|Q|} \sum_{q \in Q} \frac{1}{m_q} \sum_{k=1}^{m_q} \text{Precision}(k)
\end{equation}

\begin{equation}
    \text{NDCG} = \frac{1}{|Q|} \sum_{q \in Q} \frac{1}{\text{IDCG}_q} \sum_{k=1}^{n} \frac{2^{\text{rel}_k} - 1}{\log_2(k+1)}
\end{equation}

\begin{table}[h]
\centering
\caption{Metrics}
\label{tab:2}       
\begin{tabular}{llll}
\hline\noalign{\smallskip}
Metric & BERT & TF-IDF & Hybrid \\
\noalign{\smallskip}\hline\noalign{\smallskip}
Davies-Bouldin & 1.5238 & 1.0342 & 1.1052\\ 
Sillhouette & 0.1578 & 0.3876 & 0.3383\\ 
Intra-Cluster & 0.8819  & 0.7430 & 0.8053 \\ 
NDCG & 0.8329 & 0.7634 & \bf{0.8587}\\
MAP & 0.7858 & 0.8112 & \bf{0.8275} \\
\noalign{\smallskip}\hline
\end{tabular}   
\end{table}

BERT provides the highest intra-cluster similarity, indicating that profiles within each cluster are highly relevant to each other, thanks to its superior contextual understanding. However, BERT’s lower Silhouette and Davies-Bouldin scores suggest less distinct clustering, which may lead to profiles appearing in multiple clusters. This overlap can be advantageous for recommending diverse profiles. The NDCG score close of 0.8329 being closer to 1, indicates that BERT has strong ranking quality, meaning it successfully prioritizes profiles that closely match the intended user needs. A mAP of 0.7858 suggests that, on average, BERT achieves a high degree of precision across recommendations. 

TF-IDF shows a high Silhouette score, reflecting clear distinctions between clusters based on keyword frequency. While this leads to distinct grouping, it can be limiting when user data is expected to be inclusive among more than one clusters, potentially resulting in less accurate recommendations. Additionally, TF-IDF's lack of contextual understanding may hinder performance by missing nuanced features and ignoring text that means the same but has been expressed differently by various users. The relatively lower NDCG score than BERT aligns with the lack of contextual understanding. The mAP score proves to be higher than that of BERT reflecting ability to achieve precise clustering. 

The results of the hybrid system indicate it is the most optimal method, striking a balance between clear cluster distinctions and high contextual understanding.  This approach provides both effective separation of profiles into varied clusters and high relevance in recommendations, ensuring accuracy even with diverse data. The high NDCG score indicates that the hybrid approach successfully leverages BERT’s semantic understanding while maintaining the clear cluster separations created by TF-IDF. Further, it also achieves greater precision as suggested by mAP metrics. 

Further each of these systems were tested to recommend top 5 relevant profiles. The first step involved pulling out a target user as per Table \ref{tab:3} for whom profile recommendations would be generated. Then the recommendation functions were defined and called where the target user's data would be compared with the embeddings or vectors of the systems to return the top 5 profiles as shown in Table. 

\begin{table}[h]
\centering
\caption{Target user data}
\label{tab:3}       
\begin{tabular}{lll}
\hline\noalign{\smallskip}
Fields & Value  \\
\noalign{\smallskip}\hline\noalign{\smallskip}
Name &  Jeffery Hammond \\ 
Email &  jeffery.hammond@example.com \\
Domain  & Cybersecurity \\
Skillset & C, C++, Python, Java, HTML, CSS, SQL  \\
Cluster & 101, 97, 102 (TF-IDF,BERT,Hybrid) \\
\noalign{\smallskip}\hline
\end{tabular}
\end{table}

The variations in similarity scores and cluster assignments across the three methods highlight differences in how effectively each model captures information. Specifically, the increase in scores from TF-IDF to the Hybrid approach indicates that incorporating BERT’s contextual understanding enhances performance, providing a more accurate representation of the profiles. Overall, the algorithms demonstrated promising outcomes with their target specific user suggestions as per Table.\ref{tab:4}. 

The process was iterated for various profile data combinations   Each iteration produced satisfactory results, with relevant profiles consistently being recommended.

\begin{table*}[htbp]
\centering
\caption{Recommendations}
\label{tab:4}

\begin{subtable}{\textwidth}
\subcaption{TF-IDF Recommendations}
\begin{tabularx}{\textwidth}{|X|X|c|c|}
\hline
\textbf{Name} & \textbf{Domain/Skillset} & \textbf{Similarity score} & \textbf{Cluster} \\
\hline
Joan Evans & ai ml, C, C++, Python, Java, HTML, CSS & 0.9803 & 101 \\
Gregory Williamson & Data Mining, C, C++, Python, Java, SQL & 0.9714 & 101 \\
Alexis Moore & Java, HTML, CSS, SQL, AWS, FlutterFlow & 0.9524 & 101 \\
Joshua Hughes & Marketing,Game design, C, C++, Pytho & 0.9439 & 101 \\
Jacob Harper & SQL, AWS, Figma, Canva, Adobe XD & 0.8797 & 101 \\
\hline
\end{tabularx}
\end{subtable}

\bigskip

\begin{subtable}{\textwidth}
\subcaption{BERT Recommendations}
\begin{tabularx}{\textwidth}{|X|X|c|c|}
\hline
\textbf{Name} & \textbf{Domain/Skillset} & \textbf{Similarity score} & \textbf{Cluster} \\
\hline
Gregory Williamson & Data Mining, C, C++, Python, Java, SQL & 0.9895 & 97 \\
Alexis Moore & Java, HTML, CSS, SQL, AWS, FlutterFlow & 0.9894 & 97 \\
Joan Evans & ai ml, C, C++, Python, Java, HTML, CSS & 0.9863 & 97 \\
Joshua Hughes & Marketing,Game design, C, C++, Python & 0.9846 & 97 \\
Paula Russell & Cyber security, python, HTML, ReactJS & 0.9221 & 83 \\
\hline
\end{tabularx}
\end{subtable}

\bigskip

\begin{subtable}{\textwidth}
\subcaption{Hybrid Recommendations}
\begin{tabularx}{\textwidth}{|X|X|c|c|}
\hline
\textbf{Name} & \textbf{Domain/Skillset} & \textbf{Similarity score} & \textbf{Cluster} \\
\hline
Joan Evans & ai ml, C, C++, Python, Java, HTML, CSS & 0.9833 & 102 \\
Gregory Williamson & Data Mining, C, C++, Python, Java, SQL & 0.9804 & 102 \\
Alexis Moore & Java, HTML, CSS, SQL, AWS, FlutterFlow & 0.9709 & 102 \\
Joshua Hughes & Marketing,Game design, C, C++, Pytho & 0.9643 & 102 \\
Paula Russell & Cyber security, python, HTML, ReactJS & 0.8943 & 102 \\
\hline
\end{tabularx}
\end{subtable}

\end{table*}

\section{Shortcomings}
Despite the favourable algorithm results, there are a few shortcomings we observed during our study which we would like to highlight:

All three methods rely heavily on abundant availability of data to begin with, a challenge commonly known as the ‘cold start problem’. Without historical data, accurate profile hits were not generated. Our survey dataset served as the foundational data required to implement the algorithm, followed by the creation of
synthetic profiles that were added to the existing corpus.
When applied to incorrect and noisy data, TF-IDF establishes inaccurate relationships, as it cannot identify semantic nuances explicitly. The performance of small data is also hindered as it may fail to assess the relevance of terms, potentially resulting in unreliable recommendations. BERT is highly resource-intensive, often requiring significant computational power and memory, which leads to longer tokenization and processing times. BERT is also highly data sensitive and improper tuning can possibly affect the model's biases. As data size grows, unchecked integration of data into the hybrid model can lead to an influx of noisy information. This can result in reduced accuracy, overfitting, and diminished model interpretability.

\section{Conclusion and Future work}
\label{sec:8}
This study concluded by suggesting a unique use case and a hybrid system for profile recommendation systems that offers a promising solution to the limitations encountered with traditional methods of networking. Three recommendation models were also systematically evaluated and the hybrid approach was optimal by capturing both frequencies and contextual data, facilitating more efficient networking within academic circles. Testing on multiple target user combinations, it was demonstrated that the system was able to yield relevant recommendations. An end-to-end development setup including a mobile application was also highlighted. 
Looking ahead, the focus is on advancing the algorithm by incorporating more sophisticated and nuanced techniques to further enhance accuracy and effectiveness such as collaborative recommendation techniques. Furthermore, application will be taken to production after due optimization and be implemented within universities and various student communities.

\subsubsection{Acknowledgements}
We thank our mentor Dr. Sangeetha N whose guidance and feedback have been instrumental in shaping this research work and manuscript. This research was supported by our institution Vellore Institute of Technology, Chennai. We thank our academic department and colleagues for their insight and expertise that greatly assisted the research.

%
%
%
\bibliographystyle{splncs04}
\bibliography{export}
\end{document}